\documentclass[a4paper,onecolumn,superscriptaddress,11pt,accepted=2019-08-29]{quantumarticle}
\pdfoutput=1
\usepackage[table]{xcolor}
\usepackage{textcomp}
\usepackage[numbers,sort&compress]{natbib}
\definecolor{quantumviolet}{HTML}{53257F}
\usepackage{hyperref}
\hypersetup{citecolor=quantumviolet}
\hypersetup{colorlinks=true}
\hypersetup{linkcolor=quantumviolet}
\hypersetup{urlcolor=quantumviolet}

\usepackage{multirow}
\usepackage[utf8]{inputenc}  
\usepackage[british]{babel}  
\usepackage{graphicx} 
\usepackage[babel]{microtype}  
\usepackage{amsmath,amssymb,bm,amsfonts,mathrsfs,bbm} 
\usepackage{xspace}  
\usepackage{pgfplots}  %
\usepackage{colortbl}
\usepackage{braket}
\newtheorem{definition}{Definition}
\usepackage{graphics}

\begin{document}
	
	\title{Two-Qubit Pure Entanglement as Optimal Social Welfare Resource in Bayesian Game}

	\author{Manik Banik}
	\address{S.N. Bose National Centre for Basic Sciences, Block JD, Sector III, Salt Lake, Kolkata 700098, India.
	}
	
	\author{Some Sankar Bhattacharya}
	\address{Department of Computer Science, The University of Hong Kong, Pokfulam Road, Hong Kong.}%
	
	\author{Nirman Ganguly}
	\address{Department of Mathematics, Birla Institute of Technology and Science Pilani, Hyderabad campus, Telengana 500078,India.
	}%
	
	\author{Tamal Guha}
	\affiliation{Physics and Applied Mathematics Unit, Indian Statistical Institute, 203 B. T. Road, Kolkata 700108, India.
	}%
	
	\author{Amit Mukherjee}
	\affiliation{Optics and Quantum Information Group, The Institute of Mathematical Sciences, HBNI, CIT Campus, Taramani, Chennai 600113, India.
	}%
	
	\author{Ashutosh Rai}
	\affiliation{School of Electrical Engineering, Korea Advanced Institute of Science and Technology (KAIST), 291 Daehak-ro, Yuseong-gu, Daejeon 34141, Republic of Korea.}
	\affiliation{International Institute of Physics, Federal University of Rio Grande do Norte, 59070-405 Natal, Brazil.}
	\affiliation{Centre for Quantum Computer Science, University of Latvia, Raina Bulv. 19, Riga, LV-1586, Latvia.
	}%
	
	\author{Arup Roy}
	\affiliation{S.N. Bose National Centre for Basic Sciences, Block JD, Sector III, Salt Lake, Kolkata 700098, India.
	}%
	
	
\begin{abstract}
Entanglement is of paramount importance in quantum information theory. Its  supremacy over classical correlations has been demonstrated in a numerous information theoretic protocols. Here we study possible adequacy of quantum entanglement in Bayesian game theory, particularly in social welfare solution (SWS), a strategy which the players follow to maximize sum of their payoffs. Given a multi-partite quantum state as an advice, players can come up with several correlated strategies by performing local measurements on their parts of the quantum state. A quantum strategy is called quantum-SWS if it is advantageous over a classical equilibrium (CE) strategy in the sense that none of the players has to sacrifice their CE-payoff rather some have incentive and at the same time it maximizes sum of all players' payoffs over all possible quantum advantageous strategies. Quantum state yielding such a quantum-SWS is called a quantum social welfare advice (SWA). We show that any two-qubit pure entangled state, even if it is arbitrarily close to a product state, can serve as quantum-SWA in some Bayesian game. Our result, thus, gives cognizance to the fact that every two-qubit pure entanglement is the best resource for some operational task.
\end{abstract}
	
\maketitle
\section{Introduction}
Game theory is the study of human conflict and cooperation within a competitive situation. It has been widely used in various social and behavioral sciences, {\it e.g.}, economics \citep{Gibbons:1992}, political sciences \citep{Ordeshook:1986}, biological phenomena \citep{Colman:1995}, as well as logic, computer science, and psychology \citep{Osborne:2003}. More formally, it is a mathematical study of strategic decision making among interacting decision makers. Each decision maker is considered as a player with a set of possible actions and each one has preference over certain actions. Such preference can be modeled mathematically by associating some payoff with each of the action. First systematic study of preferences over different possible actions was discussed by von Neumann and Morgenstern \citep{Neumann:1944}. Then J. Nash introduced the seminal concept-- the concept of {\it Nash equilibrium} \citep{Nash:1950,Nash:1951}. He also proved that for any game, with finite number of actions for each player, there will always be a mixed strategy Nash equilibrium. Later, Harsanyi introduced the notion of Bayesian games where each player has some private information unknown to other players \citep{Harsanyi:1967a,Harsanyi:1967b,Harsanyi:1967c}. In such a Bayesian scenario Aumann proved that  the proper notion of equilibrium is not the ordinary mixed strategy Nash equilibrium but a more general -- {\it correlated equilibrium}  \citep{Aumann:1974}. A correlated equilibrium can be achieved by some correlated strategy where correlation is given to the players as common advice by some referee. Later it has been further established that psychology of the participating players is also an important component in the study of game theory \citep{Rabin:1991}. Psychological evidence shows that rather than pursuing solely their own payoffs, players may also consider additional social goals. Such social behavior of the players may result different types of `fairness equilibrium' solution. One such concept is {\it social welfare solution} (SWS) where the players try to maximize sum of their payoffs \citep{Binmore:1998}.
	
In this work, we study this particular notion of SWS, but in the quantum realm. In the quantum scenario the referee, instead of a classical correlation, provides a multi-partite quantum system to the players as common advice. The players can come up with correlations generated from the quantum advice by performing local measurements on their respective parts of quantum system and consequently can follow a correlated strategy. Such a quantum strategy is advantageous over a classical equilibrium (CE) strategy if none of the players' payoff is lower than the corresponding CE-payoff, rather some players have incentive over the CE-payoff. Among different advantageous quantum strategies those maximizing the sum of all players' payoffs will be  called quantum-SWS. Furthermore, a quantum state giving rise to such a strategy is called quantum social welfare advices (quantum-SWA). In this work we show that any two-qubit pure entangled state, however less entanglement it may have, can produce quantum-SWS for some Bayesian game. In other words, all such entangled states can act as useful resource for some game. We establish this claim by constructing a family of two-player Bayesian games. Rest of the paper is organized as follows. In Sec. [\ref{sec:pre}] we briefly review the framework of game theory. In Sec. [\ref{sec:adv}] we discuss some important notions regarding the use of quantum correlations as advice in games. Our main results are presented in Sec. [\ref{sec:res}], and in Sec. [\ref{sec:dis}] we present a brief discussion.

\section{Game theory: Prelude}\label{sec:pre}
\subsection{Mathematical preliminaries}
Game theory starts with a very basis assumption that the players are rational, i.e., they  will choose the best actions to get highest available payoffs \footnote{Note that situation where players have \emph{bounded rationality} is also studied in game theory \citep{March:1978,Arthur:1994}. However, in this work we will consider only rational players.}. We denote a game by the symbol $\mathbb{G}$ and for simplicity we restrict the discussion to two-player games played between (say) Alice and Bob (extension to higher number of players is straightforward and interested readers may see the classic book by Osborne \citep{Osborne:2003}). We denote the type of $i^{th}$ player by $t_i\in\mathcal{T}_i$ and denote her/his action by $s_i\in\mathcal{S}_i$, for $i\in\{A,B\}$, calligraphic fonts denoting the type and action profiles. A type can represent many things: it can be a characteristic of the player or a secret objective of the player, which remain private to the players in Bayesian scenario. There may be a prior probability distribution $P(\mathcal{T})$ over the type profile $\mathcal{T}:=\mathcal{T}_A\times\mathcal{T}_B$. Each player is assigned a payoff  over the type and action profile, i.e., $v_i:\mathcal{T}\times\mathcal{S}\mapsto \mathbb{R}$, where $\mathcal{S}:=\mathcal{S}_A\times\mathcal{S}_B$. In the absence of any correlation or external advice, players can apply strategies that are either pure or mixed. For the $i^{th}$ player, a pure strategy  is a map $g_i:\mathcal{T}_i\mapsto\mathcal{S}_i$, meaning that the player selects a deterministic action based only on her/his type. A mixed strategy is a probability distribution over pure ones, i.e. the	function $g_i:\mathcal{T}_i\mapsto\mathcal{S}_i$ becomes a random function described by a conditional probability distribution on $\mathcal{S}_i$ given the type $t_i\in\mathcal{T}_i$ and we will denote such mixed strategies as $g_i(s_i|t_i)$ (for a more detailed discussion see \citep{Auletta:2016}). The average payoff for the $i^{th}$ player is given by, $\langle v_i(g)\rangle:=\sum_{t,s}P(t)v_i(t, s)g_A(s_A|t_A)g_B(s_B|t_B)$. Here $g\equiv(g_A,g_B)\in\mathcal{G}=\mathcal{G}_A\times\mathcal{G}_B$, with $\mathcal{G}_i$ denoting the strategy profile for the $i^{th}$ party, $s\equiv(s_A,s_B)\in\mathcal{S}$, and $t\equiv(t_A,t_B)\in\mathcal{T}$; and $P(t)$ denotes the probability according to which the types are sampled. A solution for a game is a family of strategies $g\equiv(g_A,g_B)$, each for Alice and Bob respectively. A solution $g^*$ is a Nash equilibrium if no player has an incentive to change the adopted strategy, i.e., $\langle v_i(g^*)\rangle\ge\langle v_i(g_i,g^*_{\text{rest}})\rangle$, for $i\in\{A,B\}$, where $\langle v_i(g_i,g^*_{\text{rest}})\rangle$ denote the average payoff of $i^{th}$ player when all the players, but $i^{th}$ player, follow the strategy profile from $g^*$ and $i^{th}$ player follow some other strategy. 
	
In practical scenario,  achievability of Nash equilibrium is an important question. As pointed out by Aumann it can be achieved only when each of the players know other players' strategy exactly. So, he proposed a more general notion of equilibrium -- correlated Nash equilibrium \citep{Aumann:1987}. While in a mixed strategy players can choose pure strategies with probability $P(g_A,g_B)=P(g_A)P(g_B)$, with $P(g_i)$ denoting the probability distribution over the $i^{th}$ player's pure strategy, Aumann pointed out that some adviser can provide a more general probability distribution (advice) which not necessarily is in the product form. A correlated strategy is defined as the map $\mathbf{g}(\lambda)$ chosen with some probability $\lambda$ from the probability space $\Lambda$ over $\mathcal{G}=\mathcal{G}_A\times\mathcal{G}_B$. The referee chooses an element $\lambda$ from $\Lambda$ and suggests to each player $i$ to follow the strategy $g_i(\lambda)$. With such an advice from the referee, the average payoff for the $i^{th}$ player is denoted as, $\langle v_i(g(\lambda))\rangle:=\sum_{t,s,\lambda}P(t)P(\lambda)v_i(t, s)g_A(s_A|t_A,\lambda)g_B(s_B|t_B,\lambda)$.
	
A correlated strategy $g^*$ chosen with some advice $\lambda\in\Lambda$ is called a correlated Nash equilibrium if no player has an incentive while deviating from the adopted strategy. Note that, every pure/mixed Nash equilibrium is also a correlated equilibrium, however the set of correlation equilibria is strictly larger that the set of mixed strategy Nash equilibria (see Appendix-\ref{app:1}). It has also been shown that correlated equilibria are easier to compute \citep{Papadimitriou:2008}.

\subsection{Quantum game theory}
Though von Neumann is the founder father of game theory and is also a great contributor to the then nascent field of quantum mechanics, the connection between these two apparently independent fields was elusive till eighties of the last century. Blaquiere initiated the study of game theory in the domain of quantum mechanics \citep{Blaquiere1,Blaquiere2}. However, the  important development in quantum game theory occur much later after the advent of quantum information theory \citep{Meyer,Eisert1999}. Strategies in classical game theory are either pure (deterministic) or mixed (probabilistic) and no player can achieve a better payoff while shifting from the equilibrium strategy. However, Meyer in his seminal work showed that a player who implements a quantum strategy can increase the expected payoff \citep{Meyer}. The resource that is used in Meyer's formulation of quantum strategies is actually the non-classical phenomena of superposition between states. This strategy helps gaining greater payoffs than that is achievable using only classical strategies. Meyer introduced some specially designed zero-sum game. Later Eisert \textit{et al.} proposed some non-zero-sum game (Prisoner's dilemma) where the two players of the game are not in sharp opposition. But their mutual cooperation may help them gaining higher payoffs. The results of  Meyer \citep{Meyer} and Eisert et al. \citep{Eisert1999} initiated a plethora of studies on quantum game theory (see the reviews \citep{Flitney2002,Guo2008,Khan2018} and references therein). 

Recently, Brunner and Linden have initiated the study of quantum game theory in Bayesian scenario, where correlated equilibrium strategy is the relevant notion of interest \citep{Brunner:2013} (see also \citep{Cheon2008,Iqbal2015,Rai2017,Khan2019} for other related works on quantum game theory in Bayesian scenario). They have studied a \textit{cooperative} Bayesian game and shown that the classical \textit{fair} Nash equilibrium can be surpassed if quantum nonlocal correlation is provided as advice. The authors in \citep{Pappa:2015} have extended this study for \textit{conflicting} Bayesian games. The nonlocal correlations providing advantage over the classical strategies in the games studied in \citep{Brunner:2013,Pappa:2015} is obtained from the two qubit maximally entangled state. More recently, some authors of the present manuscript have shown that such nonlocal correlation turns out to be advantageous even over the \textit{unfair} correlated Nash equilibrium \citep{Roy:2016}. But surprisingly, the two-qubit maximally entangled state is not helpful here, rather, some non-maximally entangled states serve the purpose there. Here it is noteworthy that universal usefulness of quantum entanglement has been established is several tasks -- Bell game \citep{cite1} or its generalization \citep{cite4}, information processing task \citep{cite2}, channel discrimination \citep{cite3}, quantum teleportation \citep{cite5} etc. Naturally one may ask the question which quantum states exhibit the advantage over the classical resources and achieves an game theoretic equilibrium (Nash equilibrium or social welfare solution \textit{etc.}) in quantum scenario. The articles \citep{Brunner:2013,Pappa:2015} exhibit such advantage only for the maximally entangled state and the article by Roy \textit{et al.} \citep{Roy:2016} establishes the same for some specific non-maximally pure entangled states. In this present article we will show that all two qubit pure entangled states are indeed useful resource in Bayesian game theoretic scenario. To show this we will consider the concept of `social welfare solution' in quantum scenario and precisely define the concept of `quantum social welfare advice' in the following.

\section{Quantum correlations as advice}\label{sec:adv}
In quantum scenario, the referee, instead of some classical correlation, provides a bi-partite quantum state $\rho_{AB}\in\mathcal{D}(\mathbb{C}^d_A\otimes\mathbb{C}^d_B)$ as advice; $\mathcal{D}(\mathbb{C}^{d_A}_A\otimes\mathbb{C}^{d_B}_B)$ denotes the set of hermitian, positive, and trace-$1$ operators (i.e. density operator) acting on the composite Hilbert space $\mathbb{C}^{d_A}_A\otimes\mathbb{C}^{d_B}_B$. The players perform positive-operator-valued-measurements (POVM) $\{E^{x_i}_{o_i}~|~E^{x_i}_{o_i}\ge 0~\forall~o_i,x_i,~\sum_{o_i}E^{x_i}_{o_i}=\mathbb{1}_i~\forall~x_i, i\in\{A,B\}\}$, with $\mathbb{1}_i$ being the identity operator on $\mathbb{C}^{d_i}_i$, and generate an input-output probability distribution $P(\mathcal{O}_A,\mathcal{O}_B|\mathcal{X}_A,\mathcal{X}_B)\equiv\{P(o_A,o_B|x_A,x_B)~|~o_i\in\mathcal{O}_i,x_i\in\mathcal{X}_i\}$ in accordance with the Born rule, i.e., $P(o_A,o_B|x_A,x_B)=\mbox{Tr}[\rho_{AB}(E^{x_B}_{o_B}\otimes E^{x_B}_{o_B})]$. The players follow some randomized strategy according to this probability distribution. Thus a quantum strategy is specified by the triplet $\big(\rho_{AB},\{E^{x_A}_{o_A}\},\{E^{x_B}_{o_B}\}\big)$. Note that, to demonstrate an advantage over the classical correlated strategies the correlation generated from a quantum strategy need to be stronger than classical (or in other word \emph{local realistic} (LR)) correlations $\Lambda_{LR}$ (see Appendix \ref{app:2}). If the given quantum advice $\rho_{AB}$ is an entangled state \citep{Werner:1989,Horodecki:2009} then it may provide correlations which are not \emph{local-realistic}, and such correlations are commonly known as nonlocal correlations \citep{Bell:1964,Bell:1966,Brunner:2014}. In Bayesian game theoretic scenario usefulness of such nonlocal correlations over the classical correlated strategies has been demonstrated in various recent results \citep{Brunner:2013,Pappa:2015,Roy:2016}.
	
From the aforesaid discussion it is evident that to achieve a better quantum strategy (than the optimal classical strategies) the players must share entangled quantum state. More precisely, an entangled quantum advice $\rho_{AB}^{ent}$ will be called advantageous over a classical equilibrium strategy $g^*$ if the players can come up with a quantum strategy $\big(\rho^{ent}_{AB},\{E^{x_A}_{o_A}\},\{E^{x_B}_{o_B}\}\big)$ such that $\langle v_i(\rho_{AB}^{ent})\rangle\ge \langle v_i(g^*)\rangle,~\forall~i$, and strict inequality holds for some (at least one) $i$; $\langle v_i(\rho_{AB}^{ent})\rangle$ denotes the payoff for the $i^{th}$ player while following the quantum strategy $\big(\rho^{ent}_{AB},\{E^{x_A}_{o_A}\},\{E^{x_B}_{o_B}\}\big)$. 
\begin{definition}
Given a quantum advice $\rho^{ent}_{AB}$, a strategy $\big(\rho^{ent}_{AB},\{E^{x_A}_{o_A}\}^*,\{E^{x_B}_{o_B}\}^*\big)$ is optimal if no player has an incentive while deviating from the adopted strategy.
\end{definition}
The following definition will be useful to compare among different quantum advices.
\begin{definition}
A quantum advice $\rho_{AB}^{*ent}$ is called the optimal advice if there is a strategy $\big(\rho_{AB}^{*ent},\{E^{x_A}_{o_A}\}^*,\{E^{x_B}_{o_B}\}^*\big)$ such that no player has an incentive while deviating from the adopted strategy even with some other quantum advice. Such a strategy is called quantum equilibrium strategy. 
\end{definition}
The authors in \citep{Pappa:2015} have studied quantum equilibrium strategy in a conflicting Bayesian game. However, the equilibrium studied there is a fair one where players have equal payoffs. The notion of classical unfair equilibrium where different players have different payoffs, is well defined. But as noted in \citep{Roy:2016}, such a notion in quantum scenario is not pertinent, in general.
This is because, given a quantum advice $\rho^{ent}_{AB}$, there may exist more than one quantum strategies, say $\big(\rho^{ent}_{AB},\{E^{x_A}_{o_A}\}^{1^{st}},\{E^{x_B}_{o_B}\}^{1^{st}}\big)$ and $\big(\rho^{ent}_{AB},\{E^{x_A}_{o_A}\}^{2^{nd}},\{E^{x_B}_{o_B}\}^{2^{nd}}\big)$, such that both are advantageous over the classical strategy $g^*$ but Alice gets optimal payoff for $1^{st}$ strategy while Bob's payoff is optimal for $2^{nd}$ one and hence results to a conflict between the players in choosing their strategies for the given advice. In such a scenario, a relevant figure of merit for the unfair quantum strategies is social optimality solution or social welfare solution (SWS). The expected social welfare SW($g$) of a classical solution $g$ is the sum of the expected payoffs of all the players, i.e., $SW(g)=\sum_{i} \langle v_i(g)\rangle$ \citep{Binmore:1998}. Importantly, this particular notion is also relevant in social choice theory \citep{Arrow:2002,Arrow:2011}. 
\begin{definition}
Consider an classical unfair equilibrium solution $g^*$, with payoffs $\langle v_A(g^*)\rangle\ne \langle v_B(g^*)\rangle$. Among the different quantum advantageous strategies over $g^*$, a quantum strategy will be called quantum-SWS if it maximizes the sum of the payoffs. The corresponding quantum entangled state $\rho_{AB}^{ent-sw}$ producing the quantum-SWS is called quantum-social welfare advice (SWA).
\end{definition}
To say mathematically, $\rho_{AB}^{SWA}$ is a quantum-SWA if there exists some quantum strategy such that,
$\langle v_i(\rho_{AB}^{SWA})\rangle\ge\langle v_i(g^*)\rangle,~~\forall~~i$ (with strict inequality for some $i$),
and the strategy maximize $\sum_{i}\langle v_i(\rho_{AB}^{SWA})\rangle$.
In the following we will establish that all the two-qubit pure entangled states are quantum-SWA in some Bayesian game.

\section{Result} \label{sec:res}
Consider a game $\mathbb{G}(\zeta,\eta)$ played between two rational players, Alice and Bob. Each of the players has two types, i.e.,  $t_i\in \mathcal{T}_i\equiv\{0,1\}$ and two actions $s_i\in \mathcal{S}_i\equiv\{0,1\}$; $i\in\{A,B\}$. The payoffs assigned to the players depend on the respective types and actions. An utility table for the game $\mathbb{G}(\zeta,\eta)$ is given in Table-\ref{table1}.
\begin{table*}[h]
	\setlength{\tabcolsep}{1em}
	{\renewcommand{\arraystretch}{1.8}	
		\begin{tabular}{cc|c|c|c|c|l|}
			\cline{3-6}
			& & \multicolumn{2}{ |c }{$t_B=0$} & \multicolumn{2}{ |c| }{$t_B=1$} \\ \cline{3-6} 
			& & \multicolumn{1}{ |c| }{$s_B=0$} & $s_B=1$ & $s_B=0$ & $s_B=1$ \\ \cline{1-6}
			\multicolumn{1}{ |c|  }{\multirow{2}{*}{$t_A=0$} }  & $s_A=0$ & \colorbox{blue!25}{$\left(\frac{\eta \zeta+1}{4} ,\frac{\eta \zeta-1}{4}\right) $} & $\left(\frac{-2\eta +\eta \zeta+1}{4} ,\frac{-2\eta +\eta \zeta-1}{4}\right)$  & \colorbox{green!35}{$\left(\frac{2\eta+3}{4},\frac{3\eta}{4}\right) $} &  $\left(\frac{3}{4},\frac{\eta}{4}\right)$   \\ \cline{2-6}
			\multicolumn{1}{ |c|  }{} & $s_A=1$ & $\left( 0,0\right)$ & \colorbox{blue!25}{$\left(0,0 \right)$}  & $\left(\frac{3}{4},\frac{\eta}{4}\right)$ & \colorbox{green!35}{$\left(\frac{3}{4},\frac{\eta}{4}\right)$}   \\ \cline{1-6}
			\multicolumn{1}{ |c|  }{\multirow{2}{*}{$t_A=1$} } & $s_A=0$ & \colorbox{yellow!55}{$\left( \frac{-1}{4},\frac{1}{4}\right) $} & $\left( 0,0\right) $ & $\left( \frac{-\eta}{4},\frac{-2\eta+9}{4}\right) $ & \colorbox{red!35}{$\left( \frac{\eta}{4},\frac{9}{4}\right) $} \\ \cline{2-6}
			\multicolumn{1}{ |c|  }{} & $s_A=1$ & $\left(\frac{-2\eta-1}{4},\frac{-2\eta+1}{4} \right)$  & \colorbox{yellow!55}{$\left( 0,0\right) $} & \colorbox{red!35}{$\left( \frac{\eta}{4},\frac{9}{4}\right) $} & \colorbox{red!35}{$\left( \frac{\eta}{4},\frac{9}{4}\right) $} \\ \cline{1-6}
	\end{tabular}}
	\caption{(Color online) Utility table for the game $\mathbb{G}(\zeta,\eta)$ with $\zeta\in[0,2)$ and $\eta>0$. Depending on the parameters $\zeta,\eta$, the colored cells denotes different equilibria. When $1/(2-\zeta)<\eta<1/\zeta$, there are two conflicting equilibrium strategies for the type $(t_A=0,t_B=0)$, that are $(s_A=0,s_B=0)$ and $(s_A=1,s_B=1)$ (blue cells). For $\eta>1/2$ also, there are two conflicting strategies, i.e., $(s_A=0,s_B=0)$ and $(s_A=1,s_B=1)$ (yellow cells) for the type $(t_A=1,t_B=0)$.}\label{table1}
\end{table*}

From Table-\ref{table1} one can see that following are the only possible pure Nash equilibrium strategies: 

\begin{itemize}
	\item[(i)] Type $(t_A=0,t_B=0)$: in this case $(s_A=0,s_B=0)$ is an equilibrium strategy with payoff $\left((\eta \zeta+1)/4 ,(\eta \zeta-1)/4\right)$, and whenever $\eta>1/(2-\zeta)$ the strategy $(s_A=1,s_B=1)$ is also an equilibrium with payoff $\left(0 ,0\right)$. Furthermore, if the values of the parameter $\zeta$ and $\eta$ be such that $1/(2-\zeta)<\eta<1/\zeta$, then there is conflict between Alice's and Bob's preferences: Alice prefers the strategy $(s_A=0,s_B=0)$ while Bob prefers $(s_A=1,s_B=1)$.
	\item[(ii)] Type $(t_A=0,t_B=1)$: here $(s_A=0,s_B=0)$ and $(s_A=1,s_B=1)$ are two equilibria with payoffs $\left((2\eta+3)/4,3\eta/4\right)$ and $\left(3/4,\eta/4\right)$, respectively.
	\item[(iii)] Type $(t_A=1,t_B=0)$: in this case $(s_A=0,s_B=0)$ is an equilibrium with payoff $\left(-1/4,1/4\right)$, and whenever $\eta>1/2$ there is another equilibrium, that is  $(s_A=1,s_B=1)$ with payoff $\left(0 ,0\right)$. Here also the equilibrium strategies are conflicting
	\item[(iv)] Type $(t_A=1,t_B=1)$: in this case there are three equilibria $(s_A=0,s_B=1)$, $(s_A=1,s_B=0)$ and $(s_A=1,s_B=1)$ each of them having the payoff $\left(\eta/4,9/4\right)$.
\end{itemize}
Consider that the types of the players are private, i.e., unknown to other player and hence the game is Bayesian in nature. 
Each player can choose the following four pure strategies: $g^1_i(t_i)=0,~g^2_i(t_i)=1,~g^3_i(t_i)=t_i,
~g^4_i(t_i)=t_i\oplus 1$. Here $g^1_i(t_i)=0$ means that $i^{th}$ player follows the action $s_i=0$ whatever her/his type $t_i$ be, and other $g_i$'s are defined analogously where $\oplus$ denotes addition modulo $2$ operation. Altogether the players have $16$ different pure strategies $(g^l_A,g^m_B)$, with $l,m=1,2,3,4$. Straightforward calculation gives the average payoffs for these $16$ pure strategies and it turns out that classical equilibrium strategies have payoffs $\langle v_A(g*)\rangle=(3+\eta+\eta \zeta)/16$ and $\langle v_B(g*)\rangle=(9+\eta+\eta \zeta)/16$, respectively (see Appendix \ref{app:3}).
	
To establish our result, i.e, superlative behavior of all $2$-qubit pure entangled states in the above described games, first we consider the set of most general $2$-party--$2$-input--$2$-output no-signaling (NS) correlations that constitutes a polytope, say $\mathcal{P}_{NS}$. The correlations resided in $\mathcal{P}_{NS}$ have been extensively studied \citep{Barrett:2005,Pironio:2005,Brunner:2006,Dupuis:2007,Mendez:2015}. Any such correlation $P(\mathcal{O}_A,\mathcal{O}_B|\mathcal{X}_A,\mathcal{X}_B)\equiv\{P(o_A,o_B|x_A,x_B)\}\in\mathcal{P}_{NS}$, with $o_i\in\mathcal{O}_i\equiv\{+1,-1\}$ and $x_i\in\mathcal{X}_i\equiv\{0,1\}$ can be represented in a canonical form where $(P(++|00),P(+-|00),P(-+|00),P(--|00))\equiv(c_{00},m_0-c_{00},n_0-c_{00},1-m_0-n_0+c_{00})$ and the rests can be defined analogously (see Appendix \ref{app:2}). When advised by such a correlation $P\in\mathcal{P}_{NS}$, Alice's and Bob's payoffs read as:
\begin{eqnarray}\nonumber\label{eq1}
\langle v_i(P)\rangle=\frac{1}{16}\left[3^{\kappa}+\frac{\eta}{2}\left( \mathbb{B}_{CHSH}+2\zeta m_0\right) - (-1)^{\kappa}(m_0-n_0)\right],
\end{eqnarray}
with $\kappa=1$ ($\kappa=2$) for $i=A$ ($i=B$). Here, $\mathbb{B}_{CHSH}$ denotes the Bell-Clauser-Horne-Shimony-Holt (Bell-CHSH) expression, $$\mathbb{B}_{CHSH}:=\sum_{k,j=0}^1(-1)^{kj}\big\langle\langle x_A=k,x_B=j\rangle\big\rangle=4\left(\sum_{k,j=0}^1(-1)^{kj}c_{kj}-m_0-n_0+1/2\right),$$
where, $\big\langle\langle x_A,x_B\rangle\big\rangle:=\sum_{o_A,o_B=+1}^{-1}o_Ao_BP(o_A,o_B|x_A,x_B)$. Correlations that are obtainable from quantum strategies form a convex set, say $\mathcal{Q}$, which is a strict subset of the polytope $\mathcal{P}_{NS}$. As discussed earlier, a quantum strategy $\big(\rho^{ent}_{AB},\{E^{x_A}_{o_A}\},\{E^{x_B}_{o_B}\}\big)$ will be a quantum social welfare solution for the game $\mathbb{G}(\zeta,\eta)$, if $\langle v_A(P)\rangle\ge\langle v_A(g*)\rangle=(3+\eta+\eta \zeta)/16$ and $\langle v_B(P)\rangle\ge\langle v_B(g*)\rangle=(9+\eta+\eta \zeta)/16$ (with at least one the inequalities strict) and $\langle v_A(P)\rangle+\langle v_B(P)\rangle$ takes the maximum value over the set of quantum correlations. Using the expression from Eq.(\ref{eq1}), we have,
\begin{equation}\label{eq2}
\langle v_A(P)\rangle+\langle v_B(P)\rangle=\frac{1}{16}\left[12+\eta\left( \mathbb{B}_{CHSH}+2\zeta m_0\right)\right].
\end{equation}
Note that, the factor within the round brackets on the right hand side of the Eq.(\ref{eq2}), i.e., the expression $\mathbb{B}_{CHSH}+2\zeta m_0$, is actually the expression of tilted-CHSH operator studied in Ref.\citep{Acin:2012}. It has been shown in \citep{Yang:2013,Bamps:2015} that within $\mathcal{Q}$ the tilted-CHSH operator takes maximum value by a probability distribution $P(\mathcal{O}_A,\mathcal{O}_B|\mathcal{X}_A,\mathcal{X}_B)\in\mathcal{Q}$ obtained form the quantum state $|\psi\rangle_{AB}=\cos\theta|00\rangle_{AB}+\sin\theta|11\rangle_{AB}$ with the local projective measurement $E^{(x_A=0)}=\sigma_z$, $E^{(x_A=1)}=\sigma_x$ and $E^{(x_B=0)}=\cos\beta\sigma_z+\sin\beta\sigma_x$,  $E^{(x_B=1)}=\cos\beta\sigma_z-\sin\beta\sigma_x$; where $\tan\beta=\sin 2\theta$ and $\zeta=2/\sqrt{1+2\tan^2 2\theta}\in ~[0,2)$. The same choice of state and measurements also maximize the right hand side of Eq.(\ref{eq2}). This is because, if $\mathbb{B}:=\sum_{o_A,o_B,x_A,x_B} C_{o_Ao_Bx_Ax_B}P(o_A,o_B|x_A,x_B)\le \mathbb{B}_L$ is an arbitrary Bell operator with $\mathbb{B}_L$ being the local bound, then the Bell operator $\mathcal{F}_{K_1,K_2}(\mathbb{B}):=K_1\mathbb{B}+K_2$, with $K_1\in\mathbb{R}_+$ and $K_2\in\mathbb{R}$, has the local realistic bound $\mathcal{F}_{K_1,K_2}(\mathbb{B}_L)$. Moreover the points on the boundary of the set of quantum correlations that achieve the quantum maximum for $\mathbb{B}$ and $\mathcal{F}_{K_1,K_2}(\mathbb{B})$ are going to be the same. This fact also ensures that for the games where $i^{th}$ player's average payoff is of the form $\langle v_i(P\rangle)=\mathcal{F}_{K^i_1,K^i_2}(\mathbb{B})$, with some Bell operator $\mathbb{B}$ but different $K^i_j$'s for different players', the concept of unfair equilibrium fits even in the quantum regime. However this is not the case always with the game $\mathcal{G}(\zeta,\eta)$ considered in this work, and for this game the above mentioned optimal tilted-CHSH yields,

\begin{align}\nonumber
\langle v_i(P)\rangle=\frac{1}{16}\left[ 3^{\kappa} + \frac{\eta}{2}\frac{3-\cos 4\theta}{\sqrt{1+\sin^2 2\theta}}+\frac{2\eta\cos^2\theta}{\sqrt{1+2\tan^22\theta}} - (-1)^{\kappa}\frac{1}{2} \cos2\theta\left(1-\frac{1}{\sqrt{1+\sin^2 2\theta}} \right) \right].
\end{align}
	
	\begin{figure}[t!]
		\centering
		\includegraphics[height=6cm,width=8cm]{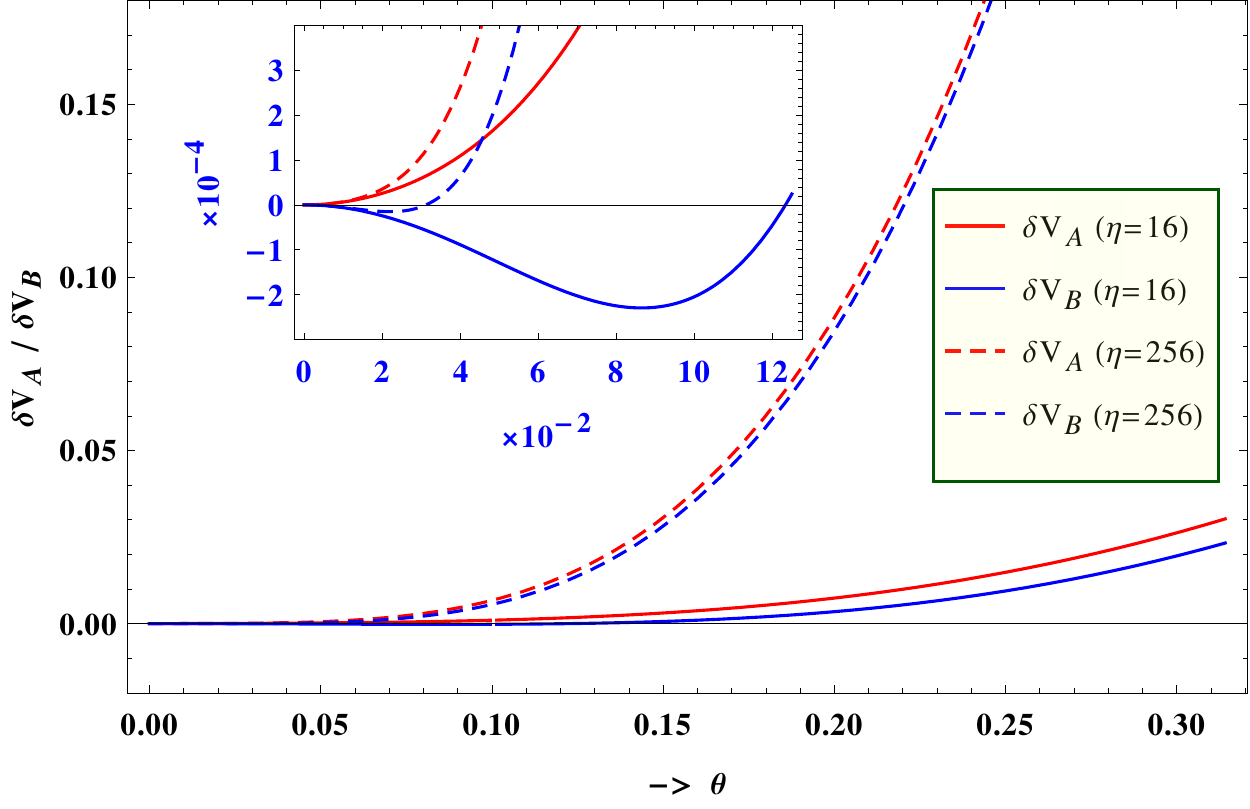}
		\caption{(Color online) $\delta V_i$ vs $\theta$ plot. Solid curves are for $\eta=16$, while dashed are for $\eta=256$. Red for $\delta V_A$ and blue for $\delta V_B$. For $eta=16$, $\delta V_B$ is positive if $\theta$ is greater than $\approx 0.12$ and for $\eta=256$ it is positive if $\theta$ is greater than $\approx 0.03$ (shown in the inset). $\delta V_A$ is positive always.}
		\label{fig}
	\end{figure}
	As already discussed, a quantum strategy will be advantageous when the players have incentive over the classical equilibrium payoff, i.e., 
	$\delta V_i:=\langle v_i(P)\rangle-\langle v_i(g^*)\rangle\ge0$ for $i\in\{A,B\}$, with strict inequality holding for at least one case. 
	Taking the value of $\eta=16$, we find that $\delta V_A>0$ for the full range of the parameter $\theta\in(0,\pi/4]$, however $\delta V_B$ remain positive if $\theta$ is not too small, if $\theta$ takes value greater than $\thickapprox 0.12$ (see Fig.\ref{fig}). Therefore the quantum states $|\psi_{AB}\rangle=\cos\theta|00\rangle+\sin\theta|11\rangle$ corresponding to the said range of $\theta$ act as the quantum social welfare advice for the game $\mathcal{G}(\zeta,\eta=16)$, where $\zeta=2/\sqrt{1+2\tan^2\theta}$. If we increase the value of $\eta$ then $\delta V_A$ remains always positive and $\delta V_B$ becomes positive for even smaller values of $\theta$ (see Fig.\ref{fig}). Moreover, taking arbitrarily large value for $\eta$ one can make $\theta$ arbitrarily close to zero and can have quantum advantage (see Appendix \ref{app:3}). It is also noteworthy that with increasing values for $\eta$ the quantum advantage over classical payoff also increases. Therefore even when the given quantum entangled state is arbitrarily close to a product state still it suffices to be a quantum-SWA.
	
	\section{Discussions}\label{sec:dis}
	Study of entanglement, its quantification, classifications as well as its applications in different information theoretic protocols \citep{Ekert:1991,Bennett:1992a,Bennett:1992b,Bennett:1993}, is one of the core research topics of quantum information theory. Quantum entanglement also draws research attention from a foundational perspective since it lies at the core of some of the most puzzling features of quantum mechanics: the Einstein-Podolski-Rosen argument \citep{Einstein:1935}, the Schrödinger’s steering concept \citep{Schrodinger:1935,Schrodinger:1936,Wiseman:2007}, and most importantly the nonlocal behavior of quantum mechanics \citep{Bell:1964,Bell:1966,Brunner:2014}. Here, we have studied an application of this quantum information theoretic resource in another vastly important area of research, Bayesian game theory. Our result establishes all two-qubit pure entangled states as the 'gold coin' in a certain Bayesian game theoretic scenario. From our analysis it is evident that the nonlocal behavior of the correlations obtained from those entangled states plays the key role in the Bayesian scenario we have considered. This observation leads us to make some interesting comments based on some already known facts. In \citep{Masanes:2006}, the authors have shown that in the $N$-party--$2$-input--$2$-output scenario the quantum maximum of any linear Bell type expression,
	$\beta:=\sum_{o_i,x_i,i\in\{1,..,N\}}C_{o_1,x_1,...,o_N,x_N}P(o_1,...,o_N|x_1,...,x_N)$,
	 is achievable by measuring $N$-qubit pure states with projective observables.     
	Therefore quantum strategies formed from these states and observables have the potential to be quantum-SWS for suitably chosen $N$-player Bayesian game where each player is given two types and two actions and where sum of the payoffs of the players turns out to be $\mathcal{F}_{K_1,K_2}(\beta)$. However, explicit construction of such games require extensive effort and promises to be an interesting topic for future research. Also note that the quantum-SWS studied in the $2-2-2$ scenario lie on the nonlocal boundary of the quantum set $\mathcal{Q}$. We leave the converse of the statement as a conjecture. We make the conjecture in a broader sense that any nonlocal boundary point of the set $\mathcal{Q}$ for general $N-M-K$ scenario is a quantum-SWS for some Bayesian game.  
	
	Another interesting question related to the present work arises from the recent interesting study of \citep{Khan2019}. The proof of Nash's theorem for the existence of an equilibrium in mixed strategies in conventional games depends on Kakutani’s fixed-point theorem \citep{Kakutani1941}. For quantum games, Meyer's study of Nash equilibrium in mixed strategies can be viewed as Glicksberg's \citep{Glicksberg1952} extension of Kakutani's fixed point theorem which does not apply directly to quantum games played with pure quantum strategies. At this point, the authors in \citep{Khan2019} made an important contribution. They have invoked Nash's famous embedding theorem \citep{Nash1956} (a more familiar result in mathematics community) and, under appropriate conditions, indirectly apply the Kakutani fixed-point theorem to guarantee Nash equilibrium in pure quantum strategies. The pure (strategy) quantum game considered in \citep{Khan2019} consists of unitary function on complex projective Hilbert space of pure quantum states. In our work we have considered correlated Nash equilibrium in the quantum scenario. It will be really interesting to make an analogous study of Ref.\citep{Khan2019} in the context of our work.
	
	\section*{Acknowledgment}
	We would like to gratefully acknowledge fruitful discussions with Guruprasad Kar. MB thankfully acknowledges discussion with Sibasish Ghosh. MB acknowledges support through an INSPIRE-faculty position at S. N. Bose National Centre for Basic Sciences, by the Department of Science and Technology, Government of India. SSB acknowledges stimulating discussions with Prof. Y C Liang. NG would like to acknowledge support from the Research Initiation Grant of BITS-Pilani, Hyderabad vide letter no. BITS/GAU/RIG/2019/H0680 dated 22nd April, 2019. Ashutosh Rai acknowledges initial support from the European Union Seventh Framework Programme (FP7/2007-2013) under the RAQUEL (Grant Agreement No. 323970) pro- ject, QALGO (Grant Agreement No. 600700) project, the ERC Advanced Grant MQC; and grant from the Brazilian ministries MEC and MCTIC. During the later phase of this work, Ashutosh Rai is supported by an Institute of Information and Communications Technology Promotion (IITP) grant funded by the Korean government (MSIP) (Grant No. 2019-0-00831, EQGIS), and ITRC Program(IITP2018-2019-0-01402). We would also like to thank the anonymous reviewers for pointing out relevant references. Their useful suggestions helped us to improve the presentation of our work.

\section*{Appendix}
\appendix
\section{Nash equilibrium}\label{app:1}
To illustrate the idea of uncorrelated and correlated Nash equilibrium, here we discuss two examples.

{\bf Example-1}: Our first example is the famous two-party game called 'battle of sexes' (BoS) where the pay-offs of the players are given as in the Table-\ref{tab1}.
\begin{table}[h!]
	\centering
	\caption{(Color Online) Utility table for the game of battle of sexes. Colored cells ($s_A=s_B$) are the two pure strategy Nash equilibria.}
	\setlength{\tabcolsep}{2em}
	{\renewcommand{\arraystretch}{1.5}
		\begin{tabular}{c|c|c|}
			\cline{2-3}
			& \multicolumn{1}{ c| }{$s_B=0$} & $s_B=1$ \\ \cline{1-3}
			\multicolumn{1}{ |c| }{$s_A=0$} & \colorbox{blue!15}{$\left( 2,1\right) $} & $\left( 0,0\right) $  \\ \cline{1-3}
			\multicolumn{1}{ |c| }{$s_A=1$} & $\left( 0,0\right) $ &\colorbox{blue!15}{$\left(1,2 \right)$}  \\ \cline{1-3}
	\end{tabular}}\label{tab1}
\end{table}
The Nash equilibria are the action profile (same as strategy profile, since the players do not have multiple types) $(s_A=0,s_B=0)$ with pay-offs $(2,1)$ and the action profile $(s_A=1,s_B=1)$ with pay-offs $(1,2)$. Now in a practical scenario, Alice and Bob can follow an equilibrium strategy if each of them deterministically know the action of other party. But if the players have ignorance about others' strategy then the achievability of equilibrium strategies are in question. In such case, a referee can advice them to reach their goal. Let the referee tosses a coin and announces the outcome (head/tail) to both Alice and Bob. Upon receiving the outcome head (tail) each party follow the strategy $s_i=0$ ($s_i=1$) and accordingly 
follow one of the equilibrium strategies. This example establishes clear practical usefulness of the idea of correlated equilibrium over the uncorrelated ones.

{\bf Example-2}: To point out more drastic difference between uncorrelated and correlated Nash equilibrium, let us consider another game known as the  'game of chickens', specified by the pay-off Table-\ref{tab2}. Here the Nash equilibria (uncorrelated) are $(s_A=0,s_B=1)$ with pay-offs $(2,7)$ and $(s_A=1,s_B=0)$ with pay-offs $(7,2)$. Also in this game there exists a uncorrelated mixed equilibrium strategy. If each player chooses the strategies $s_i=0$ and $s_i=1$ with probability $2/3$ and $1/3$, respectively then they have the equilibrium pay-off $(14/3,14/3)$. To see this, suppose player $A~(B)$ assigns probability $p~(q)$ to their respective pure action $0$. The expected payoff for $A$ $(B)$ to $s_A=0$ $(s_B=0)$ and $s_A=1$ $(s_B=1)$ are respectively $4q+2$ $(4p+2)$ and $7q$ $(7p)$. From the definition of mixed strategy equilibrium it is evident that it will be attained when each will yield the same expected payoff for both $s_i=0$ and $s_i=1$ for $i={A,B}$. This restricts both $p$ and $q$ to be $2/3$ to attain the expected payoff $(14/3,14/3)$ for the mixed strategy equilibrium.  
\begin{table}[t!]
	\centering
	\caption{(Color Online) Utility table for the game of chicken. Colored cells ($s_A\ne s_B$) are the two pure strategy Nash equilibria.}
	\setlength{\tabcolsep}{2em}
	{\renewcommand{\arraystretch}{1.5}
		\begin{tabular}{c|c|c|}
			\cline{2-3}
			& \multicolumn{1}{ c| }{$s_B=0$} & $s_B=1$ \\ \cline{1-3}
			\multicolumn{1}{ |c| }{$s_A=0$} & $\left( 6,6\right) $ & \colorbox{blue!15}{$\left( 2,7\right) $}  \\ \cline{1-3}
			\multicolumn{1}{ |c| }{$s_A=1$} & \colorbox{blue!15}{$\left( 7,2\right) $} & $\left(0,0 \right)$  \\ \cline{1-3}
	\end{tabular}}\label{tab2}
\end{table}

However like in the BoS game here also a referee can help the player to follow some particular correlated strategy. If the referee provides the players a correlation advice according to which they choose any one of pure strategies $(s_A=0,s_B=0)$, $(s_A=0,s_B=1)$, and $(s_A=1,s_B=0)$ randomly, then the average pay-off will be $(5,5)$ which is a correlated Nash equilibrium.

Note that, this correlated equilibrium can not be reached by convex mixing of the uncorrelated Nash equilibria. Clearly this shows that the notion of correlated equilibrium is more general than the original notion of equilibrium as introduced by Nash-- correlated equilibrium can be in the outside of convex hull formed by the (uncorrelated) Nash equilibrium strategies. But it is important to point out that every Nash equilibrium is a correlated equilibrium though the converse is not true. Another fundamental aspect of game theory is the degree of complexity of finding the equilibria. It was shown that correlated equilibrium are easier to be computed \citep{Papadimitriou:2008}.
\section{Correlations (as Advice)-- Local vs Nonlocal}\label{app:2}
Correlation obtained from the referee as advice helps the players to achieve the correlated equilibrium strategy. Based on different restrictions on the shared correlations, various notions of equilibrium can be defined, such as shared randomness equilibrium, no-signaling correlation equilibrium etc \citep{Auletta:2016}. On the other hand, study of correlations, in particular local vs nonlocal as inspired by the seminal result of Bell \citep{Bell:1964,Bell:1966}, is one of the fundamental aspect of quantum foundations \citep{Brunner:2014}. Very recently, Brunner and Linden have explored the connection between Bell nonlocality and Bayesian game theory \citep{Brunner:2013}. In a Bayesian game each player may have some private information unknown to other players; on the other hand, the players may have a common piece of advice and thus can follow correlated strategies. As pointed out by Brunner and Linden, the concept of private information in Bayesian games is analogous to the notion of locality in Bell inequalities (BIs), and the fact that common advice in Bayesian games does not reveal the private information mimics the concept of no-signaling resources in case of BIs. 

Correlations among spatially separated parties are relevant for our purpose. Any such correlations can be represented as input-output conditional probability distribution. Here, for our purpose, we restrict ourselves into two parties, Alice and Bob. Denoting the inputs of Alice and Bob by $x_A\in\mathcal{X}_A$ and $x_B\in\mathcal{X}_B$ and their outcomes by $o_A\in\mathcal{O}_A$ and $o_B\in\mathcal{O}_B$, the input-output probability can be represented as a conditional probability $P(\mathcal{O}_A,\mathcal{O}_B|\mathcal{X}_A,\mathcal{X}_B):=\{P(o_Ao_B|x_Ax_B)~|~o_A\in\mathcal{O}_A,o_B\in\mathcal{O}_B,x_A\in\mathcal{X}_A,x_B\in\mathcal{X}_B\}$ which must satisfy, 
\begin{enumerate}
	\item[] positivity: $P(o_A,o_B|x_A,x_B)\ge 0,~~\forall~~ o_A,o_B,x_A,x_B$, and
	\item[] normalization: $\sum\limits_{o_A,o_B}P(o_A,o_B|x_A,x_B)=1~~\forall~~ x_A,x_B$.
\end{enumerate}
Correlations compatible with the principle of `relativistic causality' principle or more generally `no signaling'(NS) principle which prevents instantaneous communication between two space-like separated locations need to satisfy further constraints:
\begin{eqnarray}
P(o_B|x_A,x_B)&:=&\sum\limits_{o_A}P(o_A,o_B|x_A,x_B)=P(o_B|x_B), \forall o_B,x_A,x_B;~~~\\\label{ns1}
P(o_A|x_A,x_B)&:=&\sum\limits_{o_B}P(o_A,o_B|x_A,x_B)=P(o_A|x_A), \forall o_A,x_A,x_B.~~\label{ns2}
\end{eqnarray} 
Any such physical correlations obtained in classical world satisfy two further conditions called \emph{locality} and \emph{reality} (LR) and are of the following form \citep{Brunner:2014}:
\begin{equation}\label{lr}
P(o_A,o_B|x_A,x_B)=\int \rho(\lambda_{LR}) P(o_A|x_A,\lambda) P(o_B|x_B,\lambda) d\lambda,
\end{equation} 
where $\lambda\in\Lambda$ is some common shared variable sampled according to the probability distribution $\rho(\lambda)$. Correlations of the form of Eq.(\ref{lr}) are also compatible with Reichenbach's principle according to which if two physical variables are found to be statistically dependent, then there should be a causal explanation of this fact \footnote{Reichenbach gave his principle a formal statement in Ref \citep{Reichenbach:1956}. In the light of Bell's theorem it's modification \citep{Cavalcanti:2014} tells that,  if two physical variables $A$ and $B$ are found to be statistically dependent the either: (i) $A$ and $B$ are directly causally connected, i.e. either $A$ causes $B$ or $B$ causes $A$, or (ii) $A$ and $B$ share a common cause that explains the correlation.}
\citep{Reichenbach:1956,Cavalcanti:2014}. However, in 1966, in the seminal paper J.S. Bell came up with an inequality \citep{Bell:1964,Bell:1966} which is satisfied by any local-realistic correlation of Eq.(\ref{lr}). Interestingly, in his paper Bell also pointed out that in quantum world correlations can arise among the outcomes of measurements performed on the entangled states of space like separated particles that violate his inequality and such are called nonlocal. 

\subsection*{$2$-party--$2$-input--$2$-output NS correlations}
Here we consider a more specific scenario with two inputs for each party with two outputs for each of the input, i.e., $o_i\in \mathcal{O}_i=\{0,1\}$ and $x_i\in \mathcal{X}_i=\{0,1\}$ for $i\in \{A,B\}$. We also consider that $\mathcal{T}_i=\mathcal{X}_i$ and $\mathcal{O}_i=\mathcal{S}_i$, that is $i$th player's types and actions correspond, respectively, to the inputs and outputs of the NS correlation. The positivity and normalization constraints for $2$-input $2$-output scenario lead the probability vector to lie in a $8$ dimensional polytope $\mathcal{P}_{NS}$ \citep{Scarani:2006}. Probability distributions satisfying the local-realistic constraint (\ref{lr}) forms another polytope $\mathcal{L}$ which is a strict subset of $\mathcal{P}_{NS}$. $\mathcal{L}$ has both trivial and nontrivial facets-- trivial facets correspond to the positivity constraints and the nontrivial ones to Bell-Clauser-Horne-Shimony-Holt (Bell-CHSH) inequality \citep{CHSH:1969}. The polytope $\mathcal{P}_{NS}$ consists of $24$ extremal points (vertices), where $16$ of them are local deterministic points being the extremal points of $\mathcal{L}$ and the rests $8$ are nonlocal extremal points. The local boxes can be written as, 
\begin{equation}
P^{\alpha,\beta,\gamma,\delta}(o_A,o_B|x_A,x_B)=
\\\begin{cases}
1, & \text{if}\ o_A=\alpha x_A \oplus \beta ~\text{and}~\ o_B=\gamma x_B \oplus \delta,  \\
~~&\\
0, & \text{otherwise},
\end{cases}
\end{equation}
with $\alpha,\beta,\gamma,\delta\in\{0,1\}$. The $16$ pure strategies $(g_A^l,g_B^m)$, with $l,m=1,2,3,4$ described in the manuscript, actually correspond to these $16$ local extremal points, i.e., the strategies are chosen according to these local deterministic extremal probability distributions.
\begin{table*}[t!]
	\centering
	\caption{(Color Online) Average pay-offs for $16$ different pure strategies for the game  $\mathbb{G}(\zeta,\eta)$.}
	\setlength{\tabcolsep}{.5em}
	{\renewcommand{\arraystretch}{1.7}
		\begin{tabular}{ |l | c | c | c | c |}
			\hline\hline
			& $g^1_B$ & $g^2_B$ & $g^3_B$ &$g^4_B~~~~~~~~~$\\ \hline \hline
			$g^1_A$ & \colorbox{blue!15}{$\left(\frac{3+\eta+\eta \zeta}{16},\frac{9+\eta+\eta \zeta}{16}\right)$} & $\left(\frac{4-\eta+\eta \zeta}{16},\frac{8-\eta+\eta \zeta}{16}\right)$ & \colorbox{blue!15}{$\left(\frac{3+\eta+\eta \zeta}{16},\frac{9+\eta+\eta \zeta}{16}\right)$} & $\left(\frac{4-\eta+\eta \zeta}{16},\frac{8-\eta+\eta \zeta}{16}\right)~~~~~~$\\ \hline
			$g^2_A$ & $\left(\frac{2-\eta}{16},\frac{10-\eta}{16}\right)$ & $\left(\frac{3+\eta}{16},\frac{9+\eta}{16}\right)$ & $\left(\frac{2-\eta}{16},\frac{10-\eta}{16}\right)$ & $\left(\frac{3+\eta}{16},\frac{9+\eta}{16}\right)~~$\\ \hline
			$g^3_A$ & \colorbox{blue!15}{$\left(\frac{3+\eta+\eta \zeta}{16},\frac{9+\eta+\eta \zeta}{16}\right)$} & $\left(\frac{4-\eta+\eta \zeta}{16},\frac{8-\eta+\eta \zeta}{16}\right)$ & $\left(\frac{3-\eta+\eta \zeta}{16},\frac{9-\eta+\eta \zeta}{16}\right)$ & $\left(\frac{4+\eta+\eta \zeta}{16},\frac{8+\eta+\eta \zeta}{16}\right)~~~~~~$\\ \hline
			$g^4_A$ & $\left(\frac{2-\eta}{16},\frac{10-\eta}{16}\right)$ & $\left(\frac{3+\eta}{16},\frac{9+\eta}{16}\right)$ & $\left(\frac{2+\eta}{16},\frac{10+\eta}{16}\right)$ & $\left(\frac{3-\eta}{16},\frac{9-\eta}{16}\right)~~$\\
			\hline
	\end{tabular}}
	\label{table2}
\end{table*}
The average payoffs for these $16$ pure strategies are calculated in Table-\ref{table2}. There are three pure strategy Nash equilibria $(g^1_A,g^1_B)$, $(g^3_A,g^1_B)$, and $(g^1_A,g^3_B)$ each average payoff $\langle v_A\rangle=(3+\eta+\eta\zeta)/16$ for Alice and average payoff $\langle v_B\rangle=(3+\eta+\eta\zeta)/16$ for Bob. Since pure/mixed strategy Nash equilibrium are also correlated equilibrium hence these are also the correlated equilibria. Moreover any convex mixture of these equilibria are again a correlated equilibria but the average payoffs for both Alice and Bob takes the same values as in the pure cases.

The advice can also be the nonlocal extremal points given by, 
\begin{equation}
P^{\alpha,\beta,\gamma}(o_A,o_B|x_A,x_B)=
\begin{cases}
1/2, & \text{if}\ o_A\oplus o_b=x_A x_B\oplus \alpha x_A \oplus \beta  x_B \oplus \gamma\\
~~&\\
0, & \text{otherwise}.
\end{cases}
\end{equation}
with $\alpha,\beta,\gamma\in\{0,1\}$, or more generally any correlation within $\mathcal{P}_{NS}$, that can be expressed as a $4\times 4$ matrix in the following canonical form:
\begin{eqnarray}\nonumber
P(\mathcal{O}_A,\mathcal{O}_B|\mathcal{X}_A,\mathcal{X}_B):=\left( \begin{array}{cccc}
c_{00} & {m_0-c_{00}} & {n_0-c_{00}} & {1-m_0-n_0+c_{00}} \\
c_{01} & {m_0-c_{01}} & {n_1-c_{01}} & {1-m_0-n_1+c_{01}} \\
c_{10} & {m_1-c_{10}} & {n_0-c_{10}} & {1-m_1-n_0+c_{10}} \\
c_{11} & {m_1-c_{11}} & {n_1-c_{11}} & {1-m_1-n_1+c_{11}}\end{array} \right),\label{NS-Cor}
\end{eqnarray}
where $(P(00|00),P(01|00),P(10|00),P(11|00))\equiv(c_{00},m_0-c_{00},n_0-c_{00},1-m_0-n_0+c_{00})$ and so on. Positivity constraint implies each element of the $4\times 4$ matrix lies in between $0$ and $1$. 

A correlation is known to be quantum one if it has a quantum realization, i.e., $P(o_A,o_B|x_A,x_B)=\mbox{Tr}[\rho_{AB}(E^{x_A}_{o_A}\otimes E^{x_B}_{o_B})]$, where $\rho_{AB}\in\mathcal{D}(\mathbb{C}^{d}_A\otimes\mathbb{C}^{d}_B)$ and $\{E^{x_A}_{o_A}\}$, $\{E^{x_B}_{o_B}\}$ represents some local POVM on Alice's and Bob's side respectively. Collection of all quantum correlations $\mathcal{Q}$ forms a convex set lying strictly in between $\mathcal{P}_{NS}$ and $\mathcal{L}$, i.e., $\mathcal{L}\subset\mathcal{Q}\subset\mathcal{P}_{NS}$. Our main interest is to study social welfare solution within the set $\mathcal{Q}$ for the the game $\mathbb{G}(\zeta,\eta)$.

\section{Pure entanglement as quantum-sw solution}\label{app:3}
If the two players are advised by a correlation from $\mathcal{P}_{NS}$ the average payoff of each player turns out to be 
\begin{eqnarray}\nonumber\label{eq01}
\langle v_A(P)\rangle&=&\frac{1}{16}\left[3+\frac{\eta}{2}\left( \mathbb{B}_{CHSH}+2\zeta m_0\right) +(m_0-n_0)\right],
\\\label{eq02}
\langle v_B(P)\rangle&=&\frac{1}{16}\left[ 9+\frac{\eta}{2}\left( \mathbb{B}_{CHSH}+2\zeta m_0\right) -m_0+n_0\right].
\end{eqnarray}
\begin{figure}
	\centering
	\includegraphics[height=8cm,width=9cm]{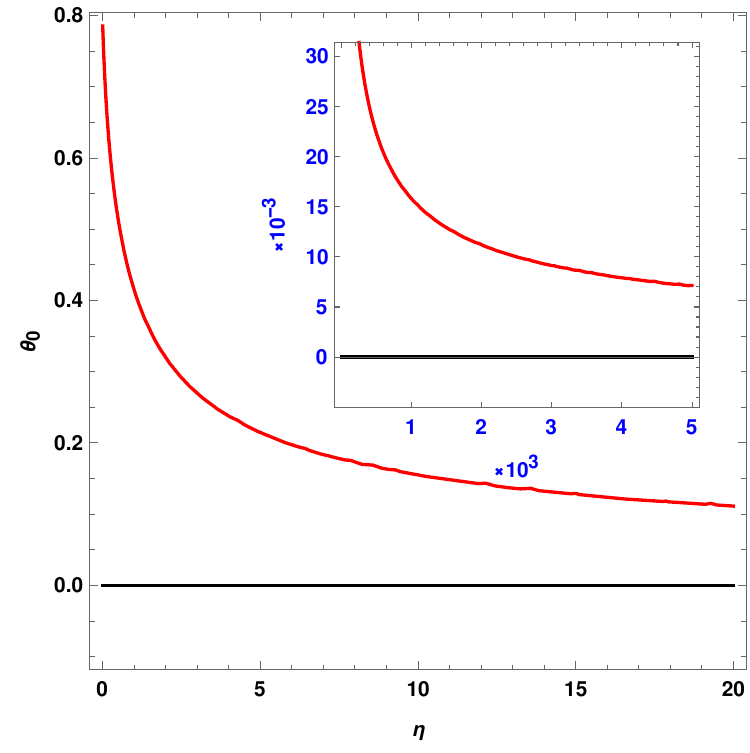}
	\caption{(Color online) $\theta_0$ vs $\eta$ plot. The graph shows that with increasing values of $\eta$ the values of $\theta_0$ gets decreased. The blue solid line is drawn for $\eta$ taking values upto $20$ and in the inset we plot it for $\eta$ upto $5000$.}
	\label{fig1}
\end{figure}

A quantum strategy will serve as a quantum social welfare solution if $\delta V_i:=\langle v_i(P)\rangle-\langle v_i(g^*)\rangle\ge0$ for $i=\{A,B\}$ and $\langle v_A(P)\rangle+\langle v_B(P)\rangle=\frac{1}{16}\left[12+{\eta}\left( \mathbb{B}_{CHSH}+2\zeta m_0\right)\right]$ yields the maximum value over $\mathcal{Q}$. 

The maximum value within $\mathcal{Q}$ of the term $\langle v_A(P)\rangle+\langle v_B(P)\rangle$ will be obtained when value of $\left( \mathbb{B}_{CHSH}+2\zeta m_0\right)$ i.e. the tilted Bell-CHSH inequality, is maximum over $\mathcal{Q}$. The above expression will reach maximum for the quantum state $|\psi\rangle_{AB}=\cos\theta|00\rangle_{AB}+\sin\theta|11\rangle_{AB}$ with the local projective measurement $E^{(x_A=0)}=\sigma_z$, $E^{(x_A=1)}=\sigma_x$ and $E^{(x_B=0)}=\cos\beta\sigma_z+\sin\beta\sigma_x$,  $E^{(x_B=1)}=\cos\beta\sigma_z-\sin\beta\sigma_x$; where $\tan\beta=\sin 2\theta$ and $\zeta=2/\sqrt{1+2\tan^2 2\theta}\in ~[0,2)$. As a result  $m_0=\cos^2\theta$, 
$n_{0}=\frac{1}{2}\left(1+\frac{\cos(2\theta)}{\sqrt{1+\sin^2(2\theta)}}\right)$ and $\mathbb{B}_{CHSH}=(3-\cos 4\theta)/\sqrt{1+\sin^2 2\theta}$, which further imply,
\begin{eqnarray}\nonumber
\langle v_A(P)\rangle=\frac{1}{16}\left[ 3 + \frac{\eta}{2}\frac{3-\cos 4\theta}{\sqrt{1+\sin^2 2\theta}} + \frac{2\eta\cos^2\theta}{\sqrt{1+2\tan^22\theta}} +\frac{1}{2} \cos2\theta\left(1-\frac{1}{\sqrt{1+\sin^2 2\theta}} \right) \right],~~~~~~~
\end{eqnarray}
\begin{eqnarray}\nonumber
\langle v_B(P)\rangle=\frac{1}{16}\left[ 9 + \frac{\eta}{2}\frac{3-\cos 4\theta}{\sqrt{1+\sin^2 2\theta}}+ \frac{2\eta\cos^2\theta}{\sqrt{1+2\tan^22\theta}} -\frac{1}{2} \cos2\theta\left(1-\frac{1}{\sqrt{1+\sin^2 2\theta}} \right) \right].~~~~~~~
\end{eqnarray}
For the classical pure equilibrium strategies $g^*\equiv\{(g^1_A,g^1_B),(g^1_A,g^3_B),(g^3_A,g^1_B)\}$, the corresponding pay-offs are,
\begin{subequations}
	\begin{align}
	\langle v_A(g^*)\rangle=\frac{3+\eta+\eta \zeta}{16}=\frac{1}{16}\left(3+\eta+\frac{2\eta}{\sqrt{1+2\tan^22\theta}} \right),\\
	\langle v_B(g^*)\rangle=\frac{9+\eta+\eta \zeta}{16}=\frac{1}{16}\left(9+\eta+\frac{2\eta}{\sqrt{1+2\tan^22\theta}} \right). 
	\end{align}
\end{subequations}
For a given $\eta$, let $\theta_0$ denotes the value of $\theta\in(0,\pi/4]$ beyond which $\delta V_B$ takes positive value. In Fig.\ref{fig1} we show how the value of $\theta_0$ tends towards zero with increasing values of $\eta$.






\end{document}